\documentclass[namedreferences]{solarphysics}
\usepackage[pdfborder={0 0 0},urlcolor=blue,breaklinks]{hyperref}
\usepackage[optionalrh]{spr-sola-addons} 
\usepackage{epsfig}                     
\usepackage{graphicx}                    
\usepackage{courier}                    
\usepackage{amssymb}                    
\usepackage{color}                       
\usepackage{url}                         

\newcommand{\etal}{{\it et al.}}
\newcommand{\arcsec}{$^{\prime\prime}$}
\newcommand{\FeX}{Fe~{\sc x}}
\newcommand{\FeXI}{Fe~{\sc xi}}
\newcommand{\FeXIII}{Fe~{\sc xiii}}
\newcommand{\FeXIV}{Fe~{\sc xiv}}
\newcommand{\kms}{km~s$^{-1}$}

\newcommand{\annG}{   {\it Ann. Geophys.}}
\newcommand{\aap}{    {\it Astron. Astrophys.}}

\newcommand{\apj}{    {\it Astrophys. J.}}
\newcommand{\apjl}{   {\it Astrophys. J. Lett.}}

\newcommand{\mnras}{  {\it Mon. Not. Roy. Astron. Soc.}}

\newcommand{\pasj}{   {\it Pub. Astron. Soc. Japan}}

\newcommand{\solphys}{{\it Solar Phys.}}

\newcommand{\ssr}{    {\it Space Sci. Rev.}}

\begin{document}

\begin{article}

\begin{opening}

\title{Variation of Emission Line Width in Mid- and High-Latitude Corona}
\author{S.~\surname{Krishna Prasad}$^{1}$\sep
        Jagdev~\surname{Singh}$^{1}$\sep
        D.~\surname{Banerjee}$^{1}$
       }
%
\runningauthor{K. Prasad \textit{et al.}}
\runningtitle{Line Width Variations with Height}
  \institute{$^{1}$ Indian Institute of Astrophysics, Bangalore-560034, India.\\
                     email: \url{krishna@iiap.res.in} email:\url{jsingh@iiap.res.in} email: \url{dipu@iiap.res.in}}          
\begin{abstract}
Spectroscopic studies of the solar corona, using the high spatial and spectral resolution 25-cm coronagraph at the Norikura observatory for equatorial off-limb observations, indicated that the variation of radiance and line width with height is different for different temperature lines. The line width of the forbidden red emission line [\FeX]~6374~\AA\ was found to increase with height and that of the green emission line [\FeXIV]~5303~\AA\ decrease with height. This had been interpreted in terms of the interaction between different temperature plasma but needed to be confirmed. Further observations were made on several days during 2004, in two emission lines simultaneously covering the mid-latitude and polar regions to investigate the existence of the observed variation in other parts of the solar corona. In this study, we have analysed several raster scans that cover mid- and high-latitude regions of the off-limb corona in all four bright emission lines [\FeX]~6374~\AA, [\FeXI]~7892~\AA, [\FeXIII]~10747~\AA, and [\FeXIV]~5303~\AA. We find that the FWHM of the red line increases with height and that of the green line decreases with height similar to that observed in equatorial regions. The line widths are found to be higher in polar regions for all of the observed emission lines except for the green line. Higher values of FWHM in polar regions may imply higher non-thermal velocities which could be further linked to a non-thermal source powering the solar-wind acceleration, but the reason for the behaviour of the green emission line remains to be explored. 
\end{abstract}

\keywords{Corona, Structures; Spectral Line, Broadening; Spectrum, Visible}
\end{opening}

\section{Introduction}
The solar corona consists of very hot, tenuous plasma, and the processes responsible for its multi-million kelvin temperature still remains unexplained. The three parameters peak radiance, Doppler shift and line width, obtained from the line profiles of coronal emission lines, provide vital information to understand the physical and dynamical properties of the source. Line profiles carry both thermal and non-thermal information of the source.

\inlinecite{1990ApJ...348L..77H} studied the line profiles of some EUV lines using a rocket experiment and observed an increased broadening with height above the limb. Coronal hole \FeX\ spectra taken at the National Solar Observatory showed an increase in the line width with height \cite{1994SSRv...70..373H}. SOHO/SUMER \cite{1995SoPh..162..189W} was used to record the off-limb, height-resolved spectra of a Si~{\sc viii} density-sensitive line pair, in  equatorial coronal regions \cite{1998SoPh..181...91D} and in polar coronal holes \cite{1998A&A...339..208B} which indicate an increase in non-thermal velocity with height. A similar analysis by \inlinecite{2004A&A...415.1133W}, in several other spectral lines, also showed an increase in Doppler width with height. Using the EIS spectrometer on {\it Hinode}, \inlinecite{2009A&A...501L..15B} reported that for  the polar region, the line width data show that the non-thermal line-of-sight velocity increases from 26~\kms\ at 10\arcsec\ above the limb to 42~\kms\ at some 150\arcsec\ ({\it i.e.} $\approx$110~000~km) above the limb. The measured variation in line width with height supports undamped wave propagation in coronal structures. This was a strong evidence of outwardly propagating, undamped Alfv\'{e}n waves in the corona, which may contribute to coronal heating and the high-speed solar wind in the case of coronal holes.  \inlinecite{1997ApJ...484L..87S}, based on SUMER observations, suggested that the increase in line width could be due to an increase in ion temperature which significantly deviates from ionization equilibrium temperature, due to preferential heating. \citeauthor{1999PASJ...51..269S} (\citeyear{1999PASJ...51..269S}, \citeyear{2002PASJ...54..807S}, \citeyear{2003ApJ...585..516S}) using extensive data obtained with the 25-cm Norikura coronagraph, found that the FWHM of [\FeX]~6374~\AA\ (red emission line) increases with height whereas that of the [\FeXIV]~5303~\AA\ (green emission line) decreases with height in the same region. They suggested that this behaviour can be explained by gradual mixing of different temperature plasma with height above the limb. By including the IR lines [\FeXIII]~10747~\AA\ and 10798~\AA\ in their study, \inlinecite{2003ApJ...585..516S} found that the FWHM of spectral lines with ionization temperatures less than 1.6~MK increase and more than 1.6~MK decrease with height. \inlinecite{2004ApJ...617L..81S} observed similar complex variations in radiance ratios of different lines, with respect to red line. The green line to red line radiance ratio decreases with height above the limb, whereas the IR to red and [\FeXI]~7892~\AA\ to red ratios increase with height. They proposed that such a behaviour in the radiance ratios of emission lines can also be explained if we consider the gradual interaction between relatively cold and hot plasma. As a further confirmation, \inlinecite{2006ApJ...639..475S} found that the FWHM of the red and green lines does not vary after heights greater than $\approx$\,250\arcsec\ above the limb. This is expected in the case of gradual mixing of multi-thermal plasma with increasing altitude, as the plasma reaches a uniform temperature and non-thermal velocity after a certain height. 

Recently, \inlinecite{2011ApJ...736..164R} reported an increase in the line width of the green line with height, up to about 400\arcsec\ above the limb, from the line profiles obtained during the total solar eclipse of 21 June 2001 using a Fabry$\--$Perot [FP] etalon. \inlinecite{2011SoPh..270..213S} also observed a case where the green-line width increases with height and showed that the FWHM of emission lines is different for different coronal features. By analysing over 130 coronal structures \inlinecite{2003ApJ...585..516S} observed that 89~\% of the structures show a negative gradient in the green emission line and 95~\% of those show a positive gradient in the red emission line. It may be just a coincidence that different structures with different physical and dynamical characteristics showed such a result reported by \inlinecite{2011ApJ...736..164R}. Also, the results by Singh \etal\ are based on a large data base spread over many years whereas those of \citeauthor{2011ApJ...736..164R} are based on observations during an eclipse of short duration. It may also be noted that the instrumental line width in their case is 0.2~\AA\ and the uncertainty in the line width determination is 0.03~\AA\ compared to the values $<$ 0.1~\AA\ and 0.002~\AA\, respectively, in the spectroscopic measurements made by \inlinecite{2002PASJ...54..793S}. In addition, the observed line profiles of emission with FP are asymmetric as compared to those with the spectrograph, because of the varying dispersion in FP interference fringes.

All of the previous studies by \citeauthor{2003ApJ...585..516S}, are mostly for equatorial regions. Here, we made an attempt to understand the behaviour of those lines in polar (high-latitude) regions, using the spectroscopic observations with the same 25-cm Norikura coronagraph. In the following sections, we give the details of our observations, discuss the data analysis steps, and finally the results obtained and their implications. 

\section{Observations}
Raster scans of the off-limb solar corona were taken using the 25-cm coronagraph at Norikura, Japan, on several days in September and October, 2004. Most of the scans were taken simultaneously in two iron emission lines using two CCD cameras, with one line being [\FeX]~6374~\AA\ (red line) and the other being [\FeXI]~7892~\AA\ or [\FeXIII]~10747~\AA\ or [\FeXIV]~5303~\AA\ (green line). A few scans were taken in the green line alone when the second camera did not work. The length of the slit is around 500\arcsec\ and the width is 4\arcsec\ for most of the scans and 5\arcsec\ for the rest. Details on the instrumental setup can be found in \citeauthor{1999PASJ...51..269S} (\citeyear{1999PASJ...51..269S}, \citeyear{2003SoPh..212..343S}, \citeyear{2003ApJ...585..516S}). Each scan covers around 100\arcsec\ of the solar corona above the limb. The exposure times were varied from scan to scan and day to day, over a broad range from 50 to 180~seconds for each spectrum, according to the target of observation to get a good signal. The dark, flat, and disk spectra were taken immediately after each set of observations. Although the main target was to scan the polar corona, due to some miscalculation during the setup, the centre of the slit was kept around 65$^{\circ}$ solar latitude which means our scans cover the region roughly from 46$^{\circ}$ to 84$^{\circ}$ solar latitude. This allowed us to study the line-width behaviour in the mid- and the high-latitude regions separately and compare them with the earlier equatorial region results.
\begin{table*}
 \caption[]{Date of observations, emission lines, pixel scale, and dispersion. Observations in a single line are blank under Line2.}
 \label{dataset}
 \begin{tabular}{c c c c c c}     
 \hline\hline
  Date      & \multicolumn{2}{c}{Emission line } & Pixel scale & \multicolumn{2}{c}{Dispersion (m\AA\ pixel$^{-1}$)} \\
 \cline{2-3}
 \cline{5-6}
               &   Line1   & Line2            & (\arcsec) & Line1   & Line2 \\
 \hline
 12 Sep. 2004  & 6374 \AA  & 7892 \AA         &  4.2      & 47.1    & 47.3 \\
 15 Sep. 2004  & 6374 \AA  & 7892 \AA         &  4.2      & 47.1    & 47.2 \\
 16 Sep. 2004  & 6374 \AA  & 7892 \AA         &  4.0      & 47.1    & 47.2 \\
 14 Oct. 2004  & 6374 \AA  & 10747 \AA        &  4.6      & 57.2    & 122.5 \\
               & 5303 \AA  & ....             &  4.0      & 46.9    & .... \\
               & 5303 \AA  & ....             &  4.0      & 46.9    & .... \\
 16 Oct. 2004  & 6374 \AA  & 10747 \AA        &  4.7      & 57.2    & 122.5 \\
               & 6374 \AA  & 10747 \AA        &  4.7      & 57.2    & 122.5 \\
 17 Oct. 2004  & 5303 \AA  & ....             &  4.0      & 39.8    & ....  \\
               & 5303 \AA  & ....             &  4.0      & 39.8    & ....  \\
               & 5303 \AA  & ....             &  4.0      & 39.8    & ....  \\
 21 Oct. 2004  & 6374 \AA  & 5303 \AA         &  4.7      & 57.2    & 31.5 \\
 25 Oct. 2004  & 6374 \AA  & 10747 \AA        &  4.7      & 57.2    & 122.0 \\
               & 6374 \AA  & 10747 \AA        &  4.7      & 57.2    & 122.1 \\
 27 Oct. 2004  & 6374 \AA  & 10747 \AA        &  4.7      & 57.2    & 123.0 \\
               & 6374 \AA  & 10747 \AA        &  4.7      & 57.2    & 123.0 \\
               & 6374 \AA  & 10747 \AA        &  4.7      & 57.2    & 123.0 \\
               & 5303 \AA  & ....             &  5.0      & 39.9    & .... \\
 \hline
 \end{tabular}
\end{table*}
Table \ref{dataset} lists the details of observations made on different dates in two emission lines simultaneously.

\section{Data Analysis}
Following the standard procedure, each spectra was corrected for the dark current, and pixel to pixel sensitivity changes using flat field and the scattered light component due to sky brightness using disk spectra. The subtraction of sky brightness, including the scattered disk light due to the instrument, yields clean emission line spectra free of absorption lines in most of the cases. The spectra obtained on some days when the exposure time was kept short due to sky conditions, were binned, both in spatial and spectral dimensions, to improve the signal-to-noise ratio. The dispersion values have been computed for each spectra and on each day of observation, using the solar-disk spectra obtained immediately after the scan and the standard solar spectra. Then the spectra is fitted with a simple Gaussian profile to obtain the parameters such as peak radiance (counts), Doppler velocity, and line width (FWHM) at each location in the solar corona. 
\begin{figure*} 
\centerline{\includegraphics[width=0.80\textwidth,clip=true]{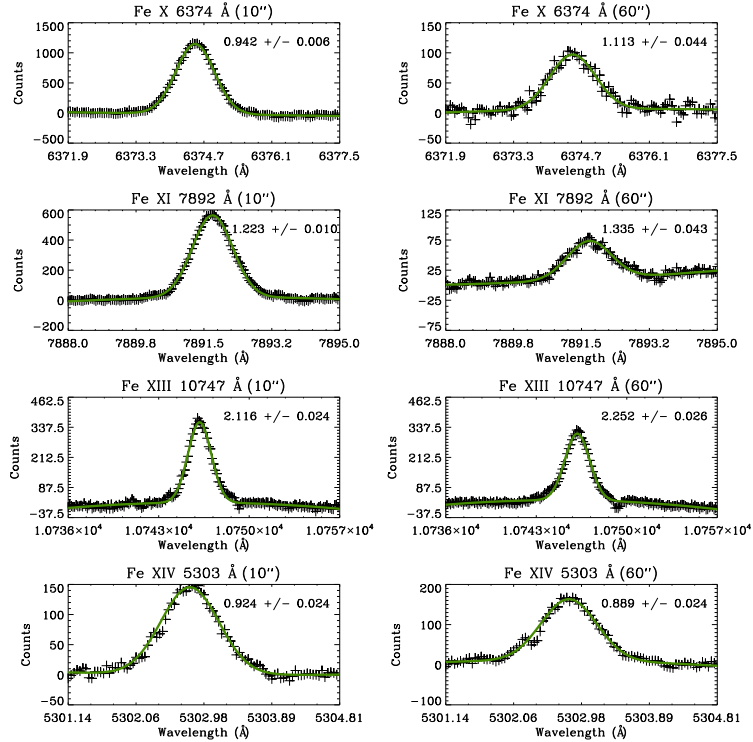}}
\caption{Line profiles of all four emission lines at two different heights 10\arcsec~(left) and 60\arcsec~(right) off-limb. The overplotted green solid lines are the fitted Gaussian profiles. Values inside each plot are the FWHM values and the corresponding uncertainties. An absorption line over the emission profile of green line (bottom row) is visible, which is discarded while fitting the profile.}
\label{lineprof}
\end{figure*}
Figure~\ref{lineprof} shows the typical line profiles in all four emission lines at two different heights (10\arcsec\ and 60\arcsec) above the limb. The overplotted solid lines (in green) are the fitted Gaussian profiles. It may be noted that the green and 7892~\AA\ emission lines have a dip (second and fourth rows from top in Figure~\ref{lineprof}) over the emission profile due to a photospheric absorption line at locations where the signal is lower. The pixel locations with this dip were carefully discarded while fitting the Gaussian profiles although the subtraction of the scattered light component from the spectrum has removed this absorption line feature in most of the spectra. The derived FWHM values from the Gaussian fit were then corrected for the instrumental effect using the relation, FWHM$_\mathrm{line}$=(FWHM$^{2}_\mathrm{obs} -$ FWHM$^{2}_\mathrm{ins}$)$^{\frac{1}{2}}$, where FWHM$_\mathrm{line}$ is the true width of the emission line, FWHM$_\mathrm{obs}$ is the observed value, and FWHM$_\mathrm{ins}$ is the instrumental contribution calculated from the disk spectra. FWHM values after this correction and the corresponding uncertainties are listed in the respective panels of Figure~\ref{lineprof}. The scans taken in two lines simultaneously have different image scales due to different focal lengths of the camera optics used. These are brought to same spatial scale using a reference slit image on the two CCD cameras taken with wires across the slit separated by a known fixed distance. The final spatial scales and the dispersion values for each scan used in this analysis are listed in Table~\ref{dataset}.

\begin{figure*} 
\centering
Radiance maps
\includegraphics[width=0.99\textwidth,clip=true]{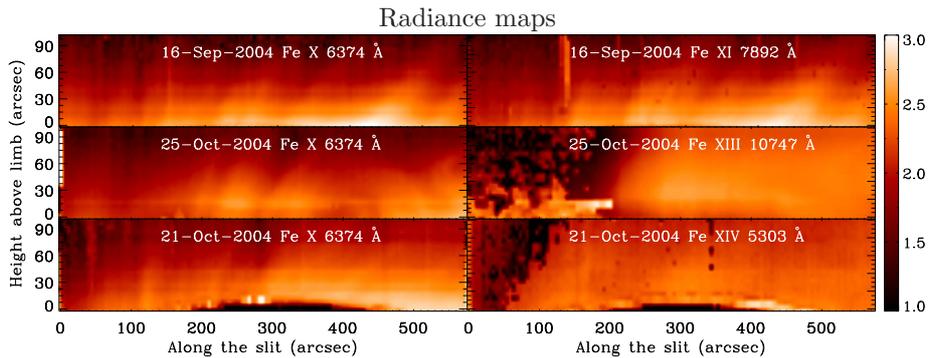}
\caption{Radiance maps (log-scaled) constructed from the raster scans taken on different dates. In each row, the left side show a radiance map of coronal region (about 500\arcsec\ $\times$ 100\arcsec), in the red emission line and the right side show corresponding region in one of the other three emission lines observed simultaneously. Respective dates and the spectral lines are written on each image.}
\label{intmaps}
\end{figure*}
Radiance and line-width maps were then constructed for each scan. A typical set of radiance maps taken simultaneously in two lines, the red line and one of the other three emission lines, can be seen in Figure~\ref{intmaps}. The radiance maps in this figure show that the coronal structures (loops) in the red and 7892~\AA\ emission lines are well defined, whereas those in the green and 10747~\AA\ lines are diffuse. EUV observations also show similar fuzziness in the coronal structures in hotter lines \cite{2009ApJ...694.1256T}. Also some portion of the coronal region can be seen to be at very low radiance, particularly in the higher-temperature emission lines. These regions correspond to the high-latitude portions of the scan.

\begin{table}
\caption{FWHM limits chosen to exclude locations with unreliable data}
\label{limits}
 \begin{tabular}{lccc}     
 \hline\hline
  Emission line     & {\it T$_{m}$}\tabnote{Temperature of maximum abundance} & \multicolumn{2}{c}{FWHM (\AA)} \\
  \cline{3-4}
  [wavelength]      & (MK)      & Lower limit & Upper limit \\
  \hline
 \FeX\ 6374~\AA\      & 1.0                 & 0.6          & 1.4 \\

 \FeXI\ 7892~\AA\     & 1.2                 & 0.8          & 1.8 \\

 \FeXIII\ 10747~\AA\  & 1.6                 & 1.3          & 2.5 \\

 \FeXIV\ 5303~\AA\    & 1.8                 & 0.6          & 1.3 \\
 \hline 
 \end{tabular}
\end{table}
We construct scatter plots from these maps to study the variation of the line width with height above the limb. At a few locations in some of the coronal regions, especially at larger heights, due to low signal-to-noise ratio, the spectra were not fitted properly leading to either very low or high FWHM values. To exclude such data locations, we restrict our analysis to the regions with FWHM values between the limits listed in Table \ref{limits}. Lower limits are calculated from the peak temperature values with zero non-thermal component and the upper limits are calculated by adding 1~MK to the peak temperature along with 30~\kms~non-thermal component (maximum reported value for these lines). However small, there will always be some non-thermal component which makes our lower limits cover the temperature regime below the peak values. 

\begin{figure*} 
\centerline{\includegraphics[width=0.98\textwidth]{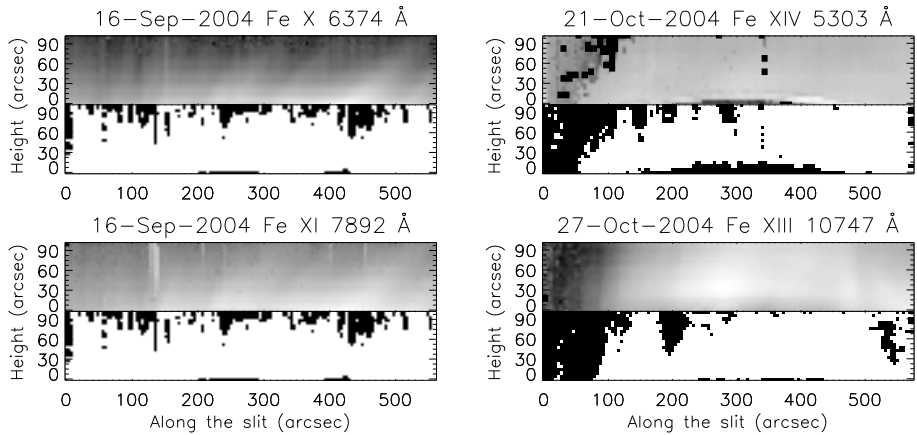}}
\caption{Radiance maps and the corresponding binary images in all four emission lines \FeX\ 6374~\AA, \FeXI\ 7892~\AA, \FeXIII\ 10747~\AA, and \FeXIV\ 5303~\AA. Radiance is displayed in the top half with the corresponding binary image in the bottom half. Dark pixel locations are the discarded locations with FWHM values outside the limits.}
\label{binmaps}
\end{figure*}
Having set the upper and lower limits for the FWHM in each emission line, we make binary images from the FWHM maps and compare them with the respective radiance maps to confirm our criterion. Binary images are constructed with 0 (dark) for locations with FWHM values outside the limits and 1 (bright) for those within limits, which are chosen for our analysis. Figure~\ref{binmaps} displays such binary images along with the corresponding radiance maps in all four emission lines. In each plot, the top half displays the radiance image and the bottom half displays the corresponding binary image. Any location with FWHM values outside the limits, in either line in the simultaneously observed pair, is discarded. The figure shows that using this criterion we exclude only a minor part of the observed coronal region from our study. The visual comparison of the radiance maps with the binary images indicates that the discarded locations with very low and high FWHM values coincides with the locations of very low radiance signal. At these locations the Gaussian fits to the observed profiles are not reliable and hence their omission in the scatter plot and our choice of limits. Moreover, our aim is to study the general variation of line width with height above the limb and any real events causing abnormally high/low line-width values should be avoided from causing contamination in our general results. 

\section{Results}
In the previous investigation over the equatorial region, \inlinecite{2006SoPh..236..245S} studied the variation in FWHM with height by choosing the locations on the loops separately and then over the diffused coronal region around loops but discarding the unreliable data of low radiance. They found that the results from the loop data, diffused plasma and the scatter plots taking the whole scan region are similar. In the current dataset, loop structures are clearly visible only in a few scans and also the simultaneous observations with green line as one of the pair are limited. Therefore, we decided to investigate the variation of FWHM with height in each line individually, using the scatter plots.

\begin{figure*}
\centering 
\includegraphics[width=0.85\textwidth]{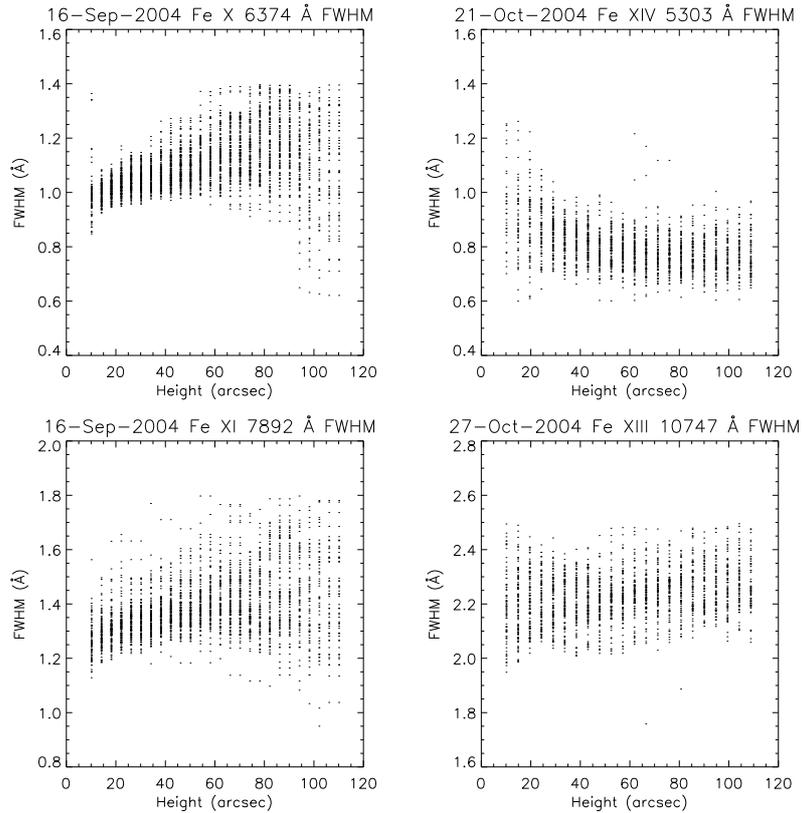}
\caption{FWHM variation over the full scan region observed, in the emission lines \FeX\ 6374~\AA, \FeXI\ 7892~\AA, \FeXIII\ 10747~\AA, and \FeXIV\ 5303~\AA. It appears that the vertical scatter at each height is due to varying physical properties of different structures along the slit.}
\label{fwhmscat}
\end{figure*}

\subsection{FWHM Variation from the Full Scan Region}
Taking the full scan region from each observation, we made scatter plots of FWHM {\it versus} height, in all four emission lines. Figure~\ref{fwhmscat} shows one such plot for each of the four emission lines observed. These plots correspond to the set of scans shown in Figure~\ref{binmaps}. In each plot, the vertical scatter at individual heights is mainly due to the varying physical characteristics of the coronal structures lying along the slit. 

\begin{figure*}
\centering 
\includegraphics[width=0.85\textwidth]{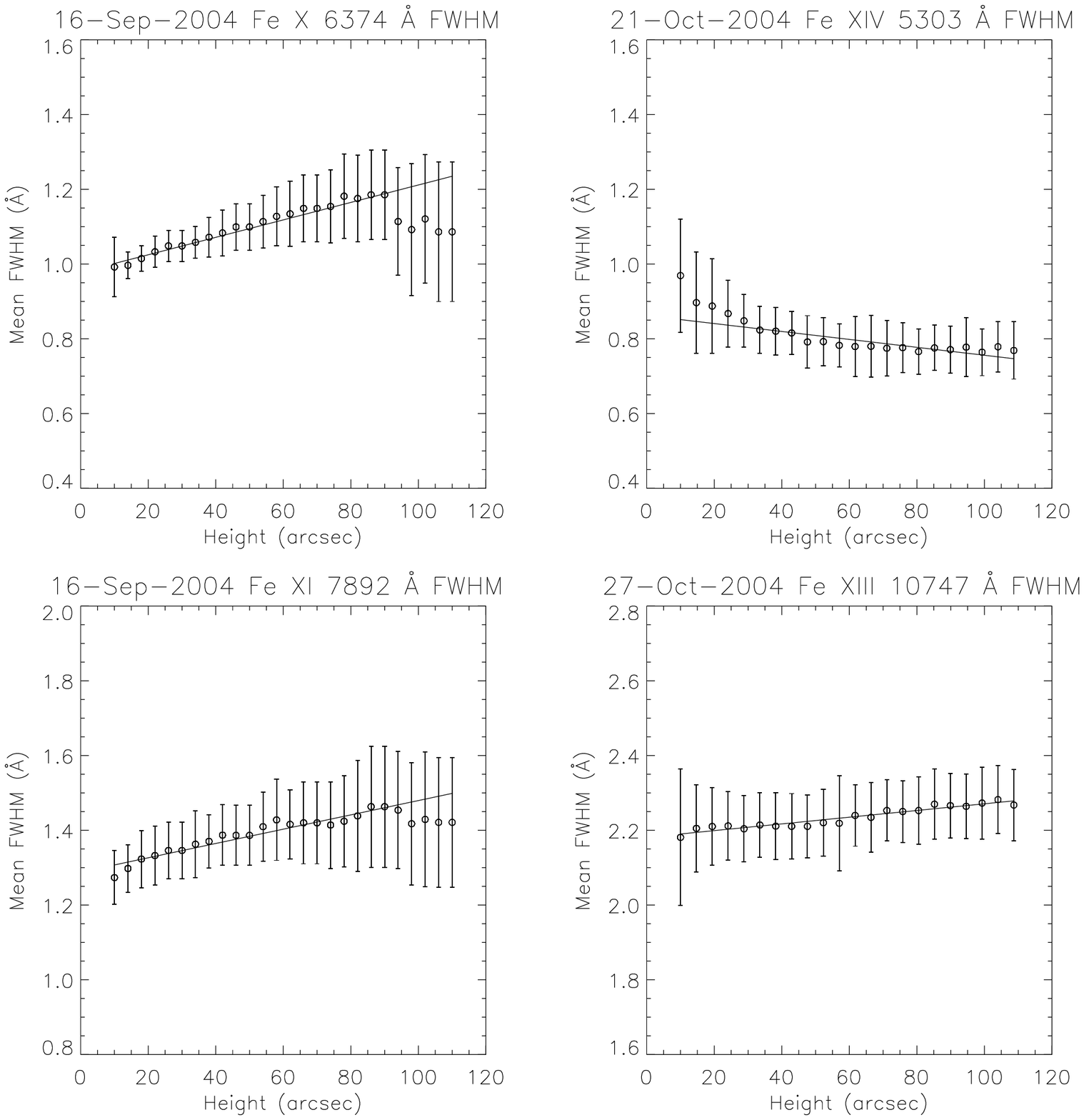}
\caption{Mean FWHM variation with height, over the full scan region observed in the emission lines \FeX\ 6374~\AA, \FeXI\ 7892~\AA, \FeXIII\ 10747~\AA, and \FeXIV\ 5303~\AA. Circles represent the mean values at individual heights and the bars denote the respective standard deviations. The solid line is the weighted linear fit to the mean values using the inverse of standard deviation as weights in fitting.}
\label{fwhmfull}
\end{figure*}

We then computed the mean values at each height and fitted them with a line to determine the gradient. Although most of the points at individual heights fall close to one another, there are some occasional outliers particularly from parts of the scan nearer and farther from the limb, which could be from the low-signal regions that crept in, despite setting the limits. To avoid this, we compute the standard deviation at each height and use them as weights in computing the linear-fit parameters. The larger is the standard deviation, the smaller is the weight given to that value. In this way, the linear fit represents most of the points and hence allows us to study the general behaviour of the observed coronal region. Figure~\ref{fwhmfull} displays the mean values and their linear fits for the plots corresponding to those in Figure~\ref{fwhmscat}. The mean values at each height are represented by circles, bars denote the corresponding standard deviations, and the solid line is the linear fit. y-axes of all the plots are scaled to the same length for better visual comparison. The linear fits show that the variation of FWHM with height is different for different-temperature lines. 

\begin{table}
\caption{FWHM mean values and their gradients over the full scans (46$^{\circ}$ to 84$^{\circ}$ latitude) for different days at 10\arcsec\ and 60\arcsec\ heights above limb, for different emission lines. Average values of all the scans and those at equatorial regions from earlier results are also listed for comparison.}
\label{comb_tbl}
\begin{tabular}{ccccc}
 \hline
Emission line & Date & \multicolumn{2}{c}{FWHM~[\AA]} & Gradient \\
\cline{3-4}
(line centre) &      & 10\arcsec & 60\arcsec & [m\AA\ arcsec$^{-1}$]\\
\hline
 6374~\AA\ & 27 Oct. 2004 & 1.07 &  1.11 &  0.90 \\
           & 25 Oct. 2004 & 1.01 &  1.07 &  1.21 \\
           & 27 Oct. 2004 & 1.07 &  1.11 &  0.77 \\
           & 14 Oct. 2004 & 1.06 &  1.10 &  0.71 \\
           & 27 Oct. 2004 & 0.97 &  1.01 &  0.85 \\
           & 16 Sep. 2004 & 0.98 &  1.13 &  2.89 \\
           & 21 Oct. 2004 & 1.02 &  1.09 &  1.43 \\
           & 15 Sep. 2004 & 1.06 &  1.07 &  0.23 \\
           & 16 Sep. 2004 & 1.00 &  1.12 &  2.33 \\
           & 25 Oct. 2004 & 1.00 &  1.04 &  0.74 \\
\hline
\multicolumn{2}{c}{Mean ($\mu\pm\sigma$)} & 1.02$\pm$0.04 & 1.08$\pm$0.04 & 1.21$\pm$0.77  \\
\multicolumn{2}{c}{Equatorial regions$^{*}$}& 0.80$\pm$0.05 & 0.86$\pm$0.06  & 1.05 \\
\hline     
  7892~\AA\ & 16 Sep. 2004 &  1.20  & 1.42 &  4.35 \\
            & 15 Sep. 2004 &  1.25  & 1.27 &  0.23 \\
            & 16 Sep. 2004 &  1.31  & 1.40 &  1.91 \\
\hline
\multicolumn{2}{c}{Mean ($\mu\pm\sigma$)} & 1.25$\pm$0.04 & 1.36$\pm$0.07 & 2.16$\pm$1.69  \\   
\multicolumn{2}{c}{Equatorial regions$^{*}$}& 1.08$\pm$0.09 & 1.10$\pm$0.10  & 0.57 \\

\hline     
  10747~\AA\ & 27 Oct. 2004 &  2.23 &  2.27 &  0.82 \\
             & 25 Oct. 2004 &  2.22 &  2.22 &  0.18 \\
             & 27 Oct. 2004 &  2.19 &  2.24 &  0.90 \\
             & 14 Oct. 2004 &  2.11 &  2.12 &  0.13 \\
             & 27 Oct. 2004 &  2.13 &  2.12 & -0.11 \\
             & 25 Oct. 2004 &  2.13 &  2.18 &  1.00 \\
\hline

\multicolumn{2}{c}{Mean ($\mu\pm\sigma$)} & 2.17$\pm$0.05 & 2.19$\pm$0.06 & 0.49$\pm$0.43  \\   
\multicolumn{2}{c}{Equatorial regions$^{*}$}& 1.86$\pm$0.10 & 1.88$\pm$0.14  & 0.22 \\

\hline     
  5303~\AA\ & 14 Oct. 2004 &  0.82 &  0.79 & -0.45 \\
            & 14 Oct. 2004 &  0.81 &  0.78 & -0.54 \\
            & 27 Oct. 2004 &  0.81 &  0.81 &  0.05 \\
            & 27 Oct. 2004 &  0.77 &  0.73 & -0.84 \\
            & 17 Oct. 2004 &  0.81 &  0.81 &  0.12 \\
            & 17 Oct. 2004 &  0.88 &  0.87 & -0.24 \\
            & 17 Oct. 2004 &  0.87 &  0.86 & -0.21 \\
            & 21 Oct. 2004 &  0.85 &  0.80 & -1.06 \\
\hline
\multicolumn{2}{c}{Mean ($\mu\pm\sigma$)} & 0.83$\pm$0.03 & 0.81$\pm$0.04 & -0.40$\pm$0.38  \\   
\multicolumn{2}{c}{Equatorial regions$^{*}$}& 0.85$\pm$0.05 & 0.82$\pm$0.06  & -0.66 \\

\hline     
\end{tabular}
 \newline
 $^{*}$Corresponding values at equatorial regions computed from the values in \\ 
Table~{\rm III} of \inlinecite{2003SoPh..212..343S}. For 10747~\AA\ values from Table~{\rm I} of \\ 
\inlinecite{2003ApJ...585..516S} are used.
\end{table}

The values of FWHM at 10\arcsec, 60\arcsec\ and the gradients obtained from the linear fits, are tabulated for all of the lines over the full scan covering the region 46$^{\circ}$ to 84$^{\circ}$ latitude, in Table~\ref{comb_tbl}. Mean values of FWHM [$\mu$], for all the observations made on different days and the standard deviation [$\sigma$] are computed and listed in a row at the end of the values for each spectral line in the table. Below the mean values, corresponding values for equatorial regions, taken from the earlier work by \citeauthor{2003SoPh..212..343S} (\citeyear{2003SoPh..212..343S}, \citeyear{2003ApJ...585..516S}), are also listed for comparison. The authors listed the FWHM values for the heights 50\arcsec\ and 100\arcsec\ above the limb and their gradients. We computed the values for heights 10\arcsec\ and 60\arcsec\ above the limb using the corresponding gradients, for direct comparison. The $\sigma$-values in gradients at equatorial regions are not listed in the table as these values are not given in the respective references. It is very important to note that the $\sigma$-values here do not represent uncertainties in determining FWHM values but indicate the dispersion in the mean values of FWHM and gradients, which arise due to different physical characteristics of different coronal structures observed on different days.

\subsubsection{Variation in FWHM of the [\FeX] Emission Line:}
The values of FWHM and the gradients (Table~\ref{comb_tbl}) indicate that the FWHM of the red line increases with height. All of the scans show a positive gradient, although the actual values vary over a broad range which might be due to different temperature or magnetic topology associated with different coronal structures. The average FWHM for these regions covering latitudes 46$^{\circ}$ to 84$^{\circ}$, is 1.02\,$\pm$\,0.04~\AA\ at the height of 10\arcsec\ as compared to 0.80\,$\pm$\,0.05~\AA\ for equatorial regions. The relative increase of this value compared with that of equatorial regions is 30~\%. The average gradient also seems to show a relative enhancement of about 15~\%.

\subsubsection{Variation in FWHM of the [\FeXI] and [\FeXIII] Emission Lines:}
The average FWHM values in 7892~\AA\ and 10747~\AA\ lines are 1.25\,$\pm$\,0.04~\AA\ and 2.17\,$\pm$\,0.05~\AA\ respectively, compared to 1.08\,$\pm$\,0.09~\AA\ and 1.86\,$\pm$\,0.1~\AA\ for the equatorial regions at 10\arcsec\ above the limb. These values indicate a relative increase of about 20~\% in the FWHM for this region over those for equatorial regions. The gradient values in the 10747~\AA\ line are mostly low and positive with negative value in one of the scans. The mean gradient value for 7892~\AA\ line seems to have changed drastically from that for equatorial regions, but the smaller data sample limits us from further analysis.

\subsubsection{Variation in FWHM of the [\FeXIV] Emission Line:}
The green emission line shows mostly negative gradients along with a couple of positive, but close to zero values, in this latitude region. The average FWHM of this line at 10\arcsec\ height above the limb is 0.83\,$\pm$\,0.03~\AA\ compared to 0.85\,$\pm$\,0.05~\AA\ for equatorial regions. The values of mean gradients are -0.40\,$\pm$\,0.38~m\AA~arcsec$^{-1}$ compared to -0.66~m\AA~arcsec$^{-1}$ for equatorial regions. It is surprising that the FWHM of the green emission line in this region is almost the same as that for equatorial region, whereas the other three emission lines show an increase in these values. It may be noted that the FWHM of the green line still decreases with height, although the gradient is less steep than in equatorial regions.

\subsection{FWHM Variation in Mid-Latitude and High-Latitude/Polar Regions:}
Each of our scans covers a broad region from 46$^{\circ}$ to 84$^{\circ}$ latitude. The results presented so far are from this full scan region over which the physical conditions can be different. So, we divided each scan into two parts one representing the mid-latitude region and the other high-latitude/polar region, by careful inspection of radiance profiles along the slit at different heights and comparing the structures with closest in epoch EIT images. This allows us to study the FWHM variations in the polar and mid-latitude regions separately and compare them. In some of the scans, particularly in polar regions, the signal is low and the reliable locations are poorly distributed. These are not suitable for the height-variation study and hence discarded. Mean FWHM values are then plotted for both mid- and high-latitude parts, and linear-fit coefficients are calculated as done before.
\begin{table}
\caption{FWHM values at 10\arcsec\ and 60\arcsec\ and their slopes for different lines in mid- and high-latitude regions. Values at equatorial regions from earlier results are also listed.}
\label{div_tbl}
\tabcolsep=0.10cm
\begin{tabular}{cccccccc}
\hline
Emission line & Date & \multicolumn{4}{c}{FWHM [\AA]} & \multicolumn{2}{c}{Gradient} \\
(line centre) &      & \multicolumn{2}{c}{10\arcsec} & \multicolumn{2}{c}{60\arcsec} & 
\multicolumn{2}{c}{[m\AA~arcsec$^{-1}$]} \\
     &    & Mid-lat. & High-lat. & Mid-lat. & High-lat. & Mid-lat. & High-lat. \\
\hline
6374~\AA\ & 27 Oct. 2004  & 1.04        & 1.09         & 1.08        & 1.17         &  0.71       & 1.66    \\
          & 25 Oct. 2004  & 1.00        & --      & 1.05        & --      &  1.11       & -- \\
          & 27 Oct. 2004  & 1.05        & 1.09         & 1.09        & 1.16         &  0.73       & 1.28    \\
          & 12 Sep. 2004  & 0.99        & --      & 1.06        & --      &  1.49       & -- \\
          & 14 Oct. 2004  & 1.05        & --      & 1.08        & --      &  0.49       & -- \\
          & 27 Oct. 2004  & 0.96        & --      & 1.01        & --      &  0.83       & -- \\
          & 21 Oct. 2004  & 0.94        & 1.12         & 1.05        & 1.14         &  2.26       & 0.49    \\
          & 15 Sep. 2004  & 1.06        & 0.96         & 1.02        & 1.12         & -0.81       & 3.13    \\
          & 16 Sep. 2004  & 1.00        & 1.00         & 1.14        & 1.10         &  2.68       & 1.93    \\
          & 25 Oct. 2004  & 1.01        & --      & 1.03        & --      &  0.44       & -- \\
  \hline
Mean ($\mu$) &  &1.01         &1.05          &1.06         &1.14          &0.99         &1.70     \\
$\pm\sigma$  &  &$\pm$0.04    &$\pm$0.06     &$\pm$0.04    &$\pm$0.03     &$\pm$0.93    &$\pm$0.86\\

\multicolumn{2}{l}{Equatorial regions$^{*}$}& \multicolumn{2}{c}{0.80$\pm$0.05} & \multicolumn{2}{c}{0.86$\pm$0.06}  & \multicolumn{2}{c}{1.05} \\
  \hline     
 7892~\AA\ & 12 Sep. 2004 & 1.22        & --      & 1.27        & --      &  0.97       & --\\
           & 15 Sep. 2004 & 1.25        & 1.15         & 1.25        & 1.26         & -0.04       & 2.20   \\
           & 16 Sep. 2004 & 1.28        & 1.34         & 1.40        & 1.40         &  2.50       & 1.28   \\
  \hline
Mean ($\mu$) &  &1.25         &1.25          &1.31         &1.33          &1.14         &1.74     \\
$\pm\sigma$  &  &$\pm$0.02    &$\pm$0.10     &$\pm$0.07    &$\pm$0.07     &$\pm$1.04    &$\pm$0.46\\

\multicolumn{2}{l}{Equatorial regions$^{*}$}& \multicolumn{2}{c}{1.08$\pm$0.09} & \multicolumn{2}{c}{1.10$\pm$0.10}    & \multicolumn{2}{c}{0.57} \\
\hline     
10747~\AA\  &  27 Oct. 2004 & 2.22        & 2.22         & 2.25        & 2.29         &  0.64       & 1.42   \\
            &  25 Oct. 2004 & 2.23        & --      & 2.21        & --      & -0.24       & --\\
            &  27 Oct. 2004 & 2.17        & 2.20         & 2.21        & 2.28         &  0.77       & 1.67   \\
            &  14 Oct. 2004 & 2.11        & --      & 2.08        & --      & -0.56       & --\\
            &  27 Oct. 2004 & 2.09        & --      & 2.09        & --      & -0.04       & --\\
            &  25 Oct. 2004 & 2.13        & --      & 2.19        & --      &  1.26       & --\\
  \hline
Mean ($\mu$) &  & 2.16         &2.21          &2.17         &2.29          &0.30         &1.54     \\
$\pm\sigma$  &  & $\pm$0.05    &$\pm$0.01     &$\pm$0.06    &$\pm$0.01     &$\pm$0.63    &$\pm$0.12\\

\multicolumn{2}{l}{Equatorial regions$^{*}$}& \multicolumn{2}{c}{1.86$\pm$0.10} & \multicolumn{2}{c}{1.88$\pm$0.14} & \multicolumn{2}{c}{0.22}\\
 \hline 
    
5303~\AA\ &  14 Oct. 2004 & 0.80        & 0.83         & 0.78        & 0.81         & -0.57       & -0.31  \\
          &  14 Oct. 2004 & 0.83        & 0.79         & 0.78        & 0.78         & -1.02       & -0.04  \\
          &  27 Oct. 2004 & 0.80        & 0.74         & 0.74        & 0.71         & -1.16       & -0.49  \\
          &  17 Oct. 2004 & 0.80        & 0.82         & 0.80        & 0.82         &  0.11       &  0.13  \\
          &  17 Oct. 2004 & 0.86        & 0.90         & 0.86        & 0.88         &  0.08       & -0.50  \\
          &  17 Oct. 2004 & 0.86        & 0.88         & 0.86        & 0.86         &  0.01       & -0.40  \\
          &  21 Oct. 2004 & 0.84        & 0.87         & 0.79        & 0.81         & -0.92       & -1.29  \\
  \hline
Mean ($\mu$) &  & 0.83         &0.83          &0.80         &0.81          &-0.50        &-0.41    \\
$\pm\sigma$  &  & $\pm$0.03    &$\pm$0.05     &$\pm$0.04    &$\pm$0.05     &$\pm$0.51    &$\pm$0.42\\

\multicolumn{2}{l}{Equatorial regions$^{*}$}& \multicolumn{2}{c}{0.85$\pm$0.05} & \multicolumn{2}{c}{0.82$\pm$0.06} & \multicolumn{2}{c}{-0.66}\\
\hline
\end{tabular}
\newline
$^{*}$Corresponding values at equatorial regions computed from the values in Table~{\rm III} of \inlinecite{2003SoPh..212..343S}. For 10747 \AA\ values from Table~{\rm I} of \inlinecite{2003ApJ...585..516S} are used.
\end{table} 
These results are tabulated in Table~\ref{div_tbl}, which indicates systematic variations in FWHM with height above the limb and also with solar latitude. FWHM values at 10\arcsec\ and 60\arcsec\ above the limb and the respective gradients both for mid- and high-latitude regions are listed in this table. Values corresponding to the discarded parts of the scans are left blank. Mean and $\sigma$ values for all of the scans, along with the values from equatorial regions, are also listed for individual lines. It can be seen that the average FWHM values in polar regions are higher than that in equatorial regions, in all of the emission lines except green line.

\begin{figure*} 
\centerline{\includegraphics[width=0.75\textwidth,clip=true]{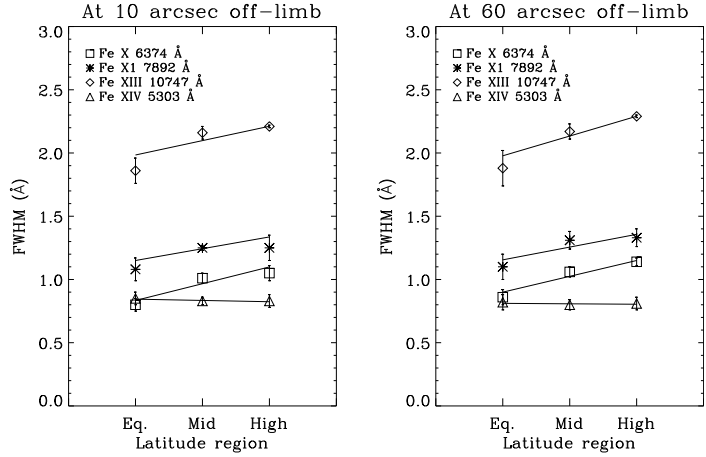}}
\caption{ Change in mean FWHM values from equatorial to mid-latitude to polar regions at 10\arcsec\ (left) and at 60\arcsec\ (right) heights above the limb in all four emission lines. Bars at each value denote the standard deviation [$\sigma$] at that value as listed in Table~\ref{div_tbl} and the solid lines are the linear fits to the mean values. Different symbols denote different emission lines as given in the legend in the plots.}
\label{latvar}
\end{figure*}

\begin{table}
\caption{Change in FWHM from equatorial to polar regions in different emission lines. Values are derived from the linear fits shown in Figure~\ref{latvar}}
\label{latchange}
 \begin{tabular}{l r@{.}l r@{.}l}     
 \hline\hline
  Emission line      & \multicolumn{4}{c}{Change in FWHM (\AA)} \\
\multicolumn{1}{l}{[wavelength]} & \multicolumn{2}{c}{at 10\arcsec} & \multicolumn{2}{l}{at 60\arcsec} \\
  \hline
  \FeX\ 6374 \AA\      & 0 & 26            & 0 & 25  \\
  \FeXI\ 7892 \AA\     & 0 & 19            & 0 & 20  \\
  \FeXIII\ 10747 \AA\  & 0 & 23            & 0 & 31  \\
  \FeXIV\ 5303 \AA\    &-0 & 02            &-0 & 01 \\
 \hline 
 \end{tabular}
\end{table}

We find a systematic change in FWHM values from equatorial to mid-latitude to high-latitude regions and it is different for different lines. To compare different lines, we plot the average FWHM values for all of the observed emission lines at the equatorial, mid-latitude, and polar regions for heights 10\arcsec\ and 60\arcsec\ above the limb in Figure~\ref{latvar}. Vertical bars over the mean values denote the dispersion of FWHM (one-$\sigma$ values) at these regions and the overplotted solid line is the linear fit to these values. It may be noted that these values do not represent a particular latitude but represent average latitudes of about 20$^{\circ}$, 55$^{\circ}$, and 75$^{\circ}$. So they are not plotted to scale in x-axis. This can alter the fit coefficients significantly, but does not affect the comparison much. Both of the panels in this figure, for 10\arcsec\ and 60\arcsec\ heights, indicate significant increase in mean FWHM from equatorial to polar regions, in all of the lines except for the green line, which shows very marginal changes. The quantitative changes for each line in these plots are given in Table~\ref{latchange}.

\section{Discussion}
Spectral line width, in the middle and high-latitude corona, is observed to vary with altitude differently in the four emission lines studied. It can be clearly seen from Figure~\ref{fwhmfull} and Tables~\ref{comb_tbl} and \ref{div_tbl} that the FWHM of the red line shows a positive gradient and that of the green line shows a negative gradient with height up to 100\arcsec\ distance off-limb. This implies that the red-line width increases with height and that of the green line decreases with height which is consistent with the earlier results for equatorial regions \cite{2003SoPh..212..343S}.  Despite the small number of scans in the other two emission lines, [\FeXI] and [\FeXIII], the trend of variation in FWHM with height in these lines, in polar regions, is similar to that observed in equatorial regions. The IR line at 10747 \AA\ shows mixed behaviour in gradient as seen in the table for mid-latitude, which is consistent with the earlier results. So this complex variation in line widths of different emission lines seems to be global and general, {\it i.e.} not restricted to any particular region or to the coronal conditions on any particular day.

Several space-based observations indicate different variations in the off-limb line widths. \inlinecite{2002A&A...392..319H} examined the Mg~{\sc x} 625 \AA\ line in the equatorial quiet region using the CDS instrument on SOHO and observed narrowing of the emission line as a function of altitude above 50~000~km. They attributed this narrowing to the dissipation of Alfv\'{e}n waves in the close field loops. However, the joint observations from CDS and SUMER by \inlinecite{2005A&A...435..733W} indicate a slight plateau rather than a decrease in the line width. Similar analysis by \inlinecite{2003A&A...400.1065O} in the north polar coronal hole, extending from the disk part of the coronal hole to $\approx$\,90~000~km above the limb suggest a turnover point, around 65~000~km above the limb, where the line widths seem to suddenly decrease or level-off. They further pointed out that this change in behaviour occurs at approximately the same line width value in each of the data sets examined, suggesting a key value for non-thermal velocity. \inlinecite{2005A&A...436L..35O} studied the height variation of both line width and line ratio of Mg~{\sc x} 609.78 and 624.94 \AA\ line pair in an off-limb north polar region. They observed that the line widths, after an initial increase, start to decrease at a height ($\approx$1150\arcsec) where the dominant excitation changes from collisional to radiative as derived from the line ratio. Based on this, it is suggested that the reduction in the non-thermal velocity, and thereby line width, is somehow linked to the excitation mechanism, probably through changes in electron density. \inlinecite{2002MNRAS.336.1195P} show that Alfv\'{e}n waves propagating along a magnetic flux tube go through refraction and get damped via viscous dissipation and resistivity under linear incompressible MHD approximation. They see a peak in the energy flux density at about 1.15~R$_\odot$ after which it declines supporting the observational evidences of damping of upwardly propagating waves. \inlinecite{2008A&A...480..509M}, using LASCO-C1 data, found a gradual increase of line width of the red line, and an initial increase followed by a decrease in the case of green line. They suggested that ion$\--$cyclotron heating or propagating Alfv\'{e}n waves can explain the gradual increase and the resonant absorption of Alfv\'{e}n waves might explain the decrease of line width after an increase. More recent observations using {\it Hinode}/EIS, reveal a gradual decrease in line width after an initial increase in polar coronal holes which is attributed to Alfv\'{e}n wave damping \cite{2012ApJ...751..110B,2012ApJ...753...36H}. All of these explanations definitely have their implications in coronal heating and solar-wind acceleration but can not explain the line width increase with height in one line and decrease in the other, of a co-spatially and simultaneously observed pair.

\inlinecite{1999PASJ...51..269S} and \inlinecite{2006ApJ...639..475S} explained this complex behaviour of line widths in terms of gradual mixing of different temperature plasma with increasing height. Although, this explanation seems to fit the observations very well, this model requires the interaction between different temperature plasma, magnetically isolated near the foot points and  gradually mixed with increase of altitude. This may not be easy in a low-$\beta$ environment such as the solar corona, more particularly in polar regions. May be the different temperature plasma regions in a loop are not magnetically isolated, instead they are physically connected through thermal conduction, as suggested by \inlinecite{2005ApJ...623..540A}. It is possible that individual loop structures that are observed have many strands at sub-pixel resolution. If we assume that these strands are fed by small-scale impulsive events such as nano flares and have shorter life span compared to the broader loop structure, then the disappearing strands with different temperature plasma may allow interaction between them. 

Now coming to the latitudinal changes, FWHM values are considerably higher in the polar region as compared to the equatorial region, for all of the lines except the green line. This behaviour of the green line is surprising. If we ignore the temperature difference between these two regions, the higher FWHM implies higher non-thermal velocities in the polar regions. These results suggest the existence of a non-thermal source, waves or turbulence, that provides additional energy to power the solar-wind acceleration and may indicate that the acceleration of the solar wind takes place at the base of the solar corona itself. But the reason for the lack of change in FWHM of the green emission line remains to be explored.

\inlinecite{2004AnGeo..22.3055C} analyzed the spectra in the green line at several off-limb locations both in equatorial and polar corona up to 12$^{\prime}$ heights above the limb. They did not find any decrease in the green-line width with height, and they observed higher values of line width in the polar regions. Our current findings do not seem to agree with their results although a direct comparison is difficult due to the observations at different coronal heights. However, our findings assume importance because of the better spectral and spatial resolutions, closely spaced observations, and the large data base generated from observations made on a number of days.

\section{Conclusions}
From our observations that cover both the mid- and high-latitude regions, we studied the dependence of FWHM on height. We compared our results with the previous results from equatorial regions obtained with the same coronagraph. We also investigated the latitudinal dependence combining the results from these three regions. We find that the FWHM of the red line shows positive gradients and that of the green line shows negative gradients with height. The gradient in the IR line is negative for some structures and positive for others. All of these results are consistent with those from equatorial regions and might indicate that this behaviour is global irrespective of the location (equatorial/polar). We also find that the FWHM values are higher in polar regions as compared to equatorial regions with a gradual increase towards the poles in all of the lines except in the green line which does not show any change. We infer that the higher FWHM values in polar regions indicate higher non-thermal velocities which might be related to some process responsible for accelerating the solar wind, but the behaviour of green line, which shows almost no change, has to be addressed. 

\acknowledgements
The authors thank the referee for useful comments. J.~Singh thanks the National Astronomical Observatory of Japan (NAOJ) for providing the financial grant to make a trip to Norikura observatory to make the observations. He also thanks the team at the observatory for their help during observations.


\end{article} 
\end{document}